\documentstyle[11pt,fleqn]{article}   
\catcode`\@=11
\@addtoreset{equation}{section}
\def\theequation{\arabic{section}.\arabic{equation}}
\def\appendix{\renewcommand{\thesection}{\Alph{section}}\setcounter{section}{0}
              \renewcommand{\theequation}
            {\mbox{\Alph{section}.\arabic{equation}}}\setcounter{equation}{0}}
\def\maketitle{\thispagestyle{empty}\setcounter{page}0\newpage
                \renewcommand{\thefootnote}{\arabic{footnote}}
                  \setcounter{footnote}0}
\renewcommand{\thanks}[1]{\renewcommand{\thefootnote}{\fnsymbol{footnote}}
               \footnote{#1}\renewcommand{\thefootnote}{\arabic{footnote}}}
\newcommand{\preprint}[1]{\hfill{\sl preprint - #1}\par\bigskip\par\rm}
\renewcommand{\title}[1]{\begin{center}\Large\bf #1\end{center}\rm\par\bigskip}
\renewcommand{\author}[1]{\begin{center}\Large #1\end{center}}
\newcommand{\address}[1]{\begin{center}\large #1\end{center}}

\def\dinfn{\smallskip Dipartimento di Fisica, Universit\`a di Trento\\ 
                           and Istituto Nazionale di Fisica Nucleare,\\
                                   Gruppo Collegato di Trento, Italia}

\def\Idinfn{\address{\dinfn}}
\newcommand{\email}[1]{e-mail: \sl #1@science.unitn.it\rm}
\newcommand{\femail}[1]{\thanks{\email{#1}}}
\newcommand{\pacs}[1]{\smallskip\noindent{\sl PACS numbers:
                       \hspace{0.3cm}#1}\par\bigskip\rm}
\def\babs{\hrule\par\begin{description}\item{Abstract: }\it} 
\def\eabs{\par\end{description}\hrule\par\medskip\rm}
\renewcommand{\date}[1]{\par\bigskip\par\sl\hfill #1\par\medskip\par\rm}
\newcommand{\ack}[1]{\par\section*{Acknowledgments} #1} 
\newcommand{\s}[1]{\section{#1}}

\renewcommand{\vec}[1]{{\bf #1}}       %%%  vectors in bold
\def\hs{\qquad}               %%%  horizontal space
\def\nn{\nonumber}            %%%  no number for eqnarray
\def\beq{\begin{eqnarray}}    %%%  begequation/eqnarray
\def\eeq{\end{eqnarray}}      %%%  endequation/eqnarray
\def\ap{\left.}               %%%  open bracket
\def\at{\left(}               %%%  open (
\def\aq{\left[}               %%%  open [
\def\ag{\left\{}              %%%  open {
\def\cp{\right.}              %%%  close bracket
\def\ct{\right)}              %%%  close )
\def\cq{\right]}              %%%  close ]
\def\cg{\right\}}             %%%  close }
\def\R{{\hbox{{\rm I}\kern-.2em\hbox{\rm R}}}}   %%% real numbers
\def\H{{\hbox{{\rm I}\kern-.2em\hbox{\rm H}}}}   %%% Hilbert space
\def\N{{\hbox{{\rm I}\kern-.2em\hbox{\rm N}}}}   %%% natural numbers
\def\C{{\ \hbox{{\rm I}\kern-.6em\hbox{\bf C}}}} %%% complex numbers
\def\Z{{\hbox{{\rm Z}\kern-.4em\hbox{\rm Z}}}}   %%% integers numbers
\def\ii{\infty}                                  %%% infinit
\def\X{\times\,}                                 %%% times
\newcommand{\fr}[2]{\mbox{$\frac{#1}{#2}$}}      %%% small fraction
\def\Tr{\mathop{\rm Tr}\nolimits}                  %%% Trace
\renewcommand{\Re}{\mathop{\rm Re}\nolimits}       %%% Real 
\renewcommand{\Im}{\mathop{\rm Im}\nolimits}       %%% Imaginary
\def\lap{\Delta}                                   %%% Laplacian
\def\al{\alpha}
\def\be{\beta}
\def\ga{\gamma}
\def\de{\delta}
\def\ep{\varepsilon}
\def\ze{\zeta}

\def\ka{\kappa}
\def\la{\lambda}

\def\om{\omega}

\def\Ga{\Gamma}

\def\La{\Lambda}

\def\citen#1{%
\edef\@tempa{\@ignspaftercomma,#1, \@end, }% ignore spaces in parameter list
\edef\@tempa{\expandafter\@ignendcommas\@tempa\@end}%
\if@filesw \immediate \write \@auxout {\string \citation {\@tempa}}\fi
\@tempcntb\m@ne \let\@h@ld\relax \let\@citea\@empty
\@for \@citeb:=\@tempa\do {\@cmpresscites}%
\@h@ld}
\def\@ignspaftercomma#1, {\ifx\@end#1\@empty\else
   #1,\expandafter\@ignspaftercomma\fi}
\def\@ignendcommas,#1,\@end{#1}
\def\@cmpresscites{%
 \expandafter\let \expandafter\@B@citeB \csname b@\@citeb \endcsname
 \ifx\@B@citeB\relax % undefined
    \@h@ld\@citea\@tempcntb\m@ne{\bf ?}%
    \@warning {Citation `\@citeb ' on page \thepage \space undefined}%
 \else%  defined
    \@tempcnta\@tempcntb \advance\@tempcnta\@ne
    \setbox\z@\hbox\bgroup % check if citation is a number:
    \ifnum\z@<0\@B@citeB \relax
       \egroup \@tempcntb\@B@citeB \relax
       \else \egroup \@tempcntb\m@ne \fi
    \ifnum\@tempcnta=\@tempcntb % Number follows previous--hold on to it
       \ifx\@h@ld\relax % first pair of successives
          \edef \@h@ld{\@citea\@B@citeB}%
       \else % compressible list of successives
          \edef\@h@ld{\hbox{--}\penalty\@highpenalty \@B@citeB}%
       \fi
    \else   %  non-successor--dump what's held and do this one
       \@h@ld \@citea \@B@citeB \let\@h@ld\relax
 \fi\fi%
 \let\@citea\@citepunct
}
\def\@citepunct{,\penalty\@highpenalty\hskip.13em plus.1em minus.1em}%
\def\@citex[#1]#2{\@cite{\citen{#2}}{#1}}%
\def\@cite#1#2{\leavevmode\unskip
  \ifnum\lastpenalty=\z@ \penalty\@highpenalty \fi % highpenalty before
  \ [{\multiply\@highpenalty 3 #1% % triple-highpenalties within list
      \if@tempswa,\penalty\@highpenalty\ #2\fi % and before note.
    }]\spacefactor\@m}
\begin{document}
%\tableofcontents       %%%%%%   index of section

\preprint{UTF 377} 
\title{
Quantum Fields in Hyperbolic Space-Times \\ 
with Finite Spatial Volume}
\author{Andrei A. Bytsenko\thanks{email: abyts@spin.hop.stu.neva.ru}}
\address{State Technical University, St. Petersburg 195251, Russia}
\author{Guido Cognola\femail{cognola} and 
Sergio Zerbini\femail{zerbini}}
\Idinfn

\date{May 1996}

\babs 
The one-loop effective action for a massive  self-interacting scalar 
field is investigated  in $4$-dimensional 
ultrastatic space-time $ R \times H^3/\Ga $, $ H^3/\Ga$ being a non-compact 
hyperbolic manifold with finite volume.  
Making use of the Selberg trace formula, 
the $\zeta$-function related to the small disturbance 
operator is constructed. For an arbitrary gravitational coupling, it is found that   
$\zeta(s)$ has a simple pole at $s=0$. 
The one-loop effective action is 
analysed by means of proper-time regularisations and 
the one-loop divergences are explicitly found. It is pointed out that, 
in this special case, 
also $\zeta$-function regularisation requires a divergent counterterm, 
which however is not necessary in the  free massless conformal invariant coupling 
 case.
Finite temperature effects are studied and the high-temperature 
expansion is presented. A possible application to the 
problem of the divergences 
of the entanglement entropy for a free massless scalar field in a 
Rindler-like space-time is briefly discussed.
\eabs


\s{Introduction}

In the last decades there has been a lot of investigations about the 
properties of quantum fields in curved space-times 
\cite{birr82b,full89b}. There has  also been some interest in 
studying free scalar fields in static  topologically nontrivial 
space-times. We would like to remind that a lot of investigations have 
been concerned with space-times with the toroidal topology 
(see for example Refs.~\cite{camp90-196-1,eliz94b} 
and references therein) or orbifold factors of 
spheres \cite{kenn81-23-2884,chan93-395-407}. The case of compact  
hyperbolic manifolds has also  been considered 
(see for example Refs.~\cite{eliz94b,byts96-266-1}). 
In this case one is dealing with an ultrastatic 
space-time $ R\times H^3/\Ga$, $H^3/\Ga$ being a compact hyperbolic 
manifold, $H^3$ the Lobachevsky space and $\Ga$  a 
discrete group of isometries acting on  $H^3$ and containing 
loxodromic, hyperbolic and elliptic elements (see 
Refs.~\cite{venk90b,elst85-17-83,elst87-277-655,byts92-33-3108,byts96-266-1}).

In this paper we extend the analysis considering a self-interacting scalar 
field in an ultrastatic space-times with  non-compact, but with finite 
volume hyperbolic spatial  section. Such spaces, for a fixed 
value of the cosmological time, may be of significant interest in 
cosmology \cite{elli71-2-7}, since it has been recently claimed that 
hyperbolic compact space-times have some problem with regard to 
inflationary scenario \cite{ishi96u-330}. In our example, 
the discrete group of isometry can be chosen in the form    
$\Ga=SL(2,\Z+i\Z)/\{\pm Id\}$. It is generated by parabolic mappings (Id 
is an isolated identity element of $\Ga$) and is associated with a 
non-compact manifold having 
an invariant fundamental domain of finite volume. In fact there exists 
a homeomorphism between  $H^3/\Ga$ and the manifold without boundary 
$(S^1 \times U_c)/\partial M$, $U_c$ being a punctured cylinder 
and the boundary $\partial M$ being homeomorphic to the torus $S^1 \times S^1$.

We shall work in the one-loop approximation. 
Making use of the Selberg trace formula, we will investigate the 
asymptotic expansion of the heat-kernel trace 
$\Tr\exp(-tL_4)$, $L_4$ 
being the small disturbance (Laplace-like) operator. We shall find that 
the presence of parabolic elements in $\Ga$ leads to the appearance of 
logarithmic factor in the small $t$ asymptotic expansion. For non-compact 
Riemannian surfaces of finite area, this fact has been observed in Refs. 
\cite{efra88-119-443,mull92-109-265}. 
However in this case the meromorphic 
continuation of the $\zeta$-function has been shown to be regular at
$s=0$, thus the determinant of the Laplacian has been evaluated by 
means of the standard $\zeta$-function regularisation 
\cite{ray71-7-145,hawk77-55-133}. On the contrary, in the
(1+3)-dimensional case, we shall show that $\zeta(s|L_4)$
is a  meromorphic function having a simple pole  at $s=0$. 
In a different context, i.e. the 
computation of functionals determinant of Laplacian 
on the generalised cone, the 
appearance of a non standard logarithmic term 
has been recently pointed out  in 
Ref.~\cite{bord96u-89} and this fact was first noted by Cheeger 
\cite{chee83-18-575} and others authors 
\cite{call83-88-357,brun85-58-133}. 
However, in Ref.~\cite{bord96u-89} the things 
have been arranged in order to avoid the logarithmic term, by dealing 
with a conformally coupled free massless field. We will show that also 
in our case, the same choice leads to a regular $\ze$-function at 
$s=0$.    

In order to give a meaning to the determinant of the 
small disturbance operator, we shall make use of the proper-time 
Schwinger regularisation, which includes the  $\zeta$-function one. 
Within this approach, the one-loop ultraviolet divergences  
will be  explicitly calculated. 
It will be found that  also the  regularisation 
requires  a counterterm, which, however, is absent for a 
conformally coupled free massless scalar field. 
Besides the usual ones we find new divergences, 
which, however, may be absorbed making use 
of the one-loop renormalisation procedure.    

The finite temperature effects as well as the vacuum 
energy will be studied in some detail for a self-interacting scalar field. 
In absence of parabolic element, these effects have been investigated 
in Ref. \cite{cogn94-49-5307}. We find that, in the high-temperature expansion, the parabolic 
elements induce  new terms, which may play a role in the cosmological applications.
    
The contents of the paper are the following.
In Sec.~2 the trace of heat kernel for a Laplace-type  operator
is computed by means of the Selberg trace formula. The meromorphic 
structure of the related $\zeta$-function is determined. 
In Sec.~3 the regularisation of the one-loop effective action
is discussed in the framework of the Schwinger regularisations. 
In Sec.~4 the finite temperature effects are investigated. 
Finally we end with some conclusions in Sec.~5.
In the Appendices, a summary of the Selberg trace formula for a non-compact
hyperbolic manifold with finite invariant fundamental domain is presented.

\s{The heat kernel and the $\zeta$-function} \label{S:HK}

Here we concentrate our attention on a self-interacting scalar field
in an ultrastatic space-time $\cal M$ in which the spatial
section is a manifold with constant curvature. For the sake of
generality, we shall derive a general expression for the one-loop
effective potential. Then we shall
discuss in detail the case in which the spatial section is
$H^3/\Gamma$, the discrete group of isometries $\Ga$ containing parabolic
elements only. When $\Gamma$ contains elliptic elements, conical-like
singularities appear in the manifold and in this case the
renormalisation becomes quite delicate. For a discussion of similar
situations see for example Ref.~\cite{cogn94-49-1029}.

We start with the classical Euclidean action for the scalar field in 
$\R \times H^3/\Gamma$ 
\begin{eqnarray} S[\phi,g]=\int_{\cal M}
\left[-\frac12\phi\Delta\phi +V_c(\phi)\right]\sqrt{g}d^4x
\:,\label{Sc}\end{eqnarray} 
where the classical potential reads
\begin{eqnarray} V_c(\phi)=\frac{\lambda\phi^4}{24}
+\frac{m^2\phi^2}2+\frac{\xi R\phi^2}2 \:,\nonumber\end{eqnarray}
$\lap$ being the Laplace-Beltrami operator,
$R$ the scalar curvature, $m$ the mass, $\la$ and $\xi$ arbitrary 
constants. 
The small disturbance operator associated with the action ~(\ref{Sc})
is given by
\begin{eqnarray} 
L_4=-\lap_4+V_c''(\phi_c)
=-\lap_4+\al^2\:,\nonumber\end{eqnarray} 
where we assume $\al^2=m^2+\xi R+\lambda\phi_c^2/2$ to be a 
constant and the prime
stands for the derivative with respect to the classical field $\phi_c$.
Since we are dealing with an ultrastatic space-time, we have
\beq
L_4=-\partial^2_\tau-\lap_3+\al^2=-\partial^2_\tau+L_3
\:,\label{nmop}\eeq
$\lap_3$ being the Laplace-Beltrami operator acting in $H^3/\Ga$.

The one-loop effective potential is defined by
\beq
V_{eff}(\phi_c)=V(\phi_c)+\frac{1}{2\ell V(F)}\ln\det\fr{L_4}{\mu^2}
\:,\label{ep}\eeq
where $\mu$ is a dimensional renormalisation parameter, $\ell $ 
the (infinite) interval of imaginary time and $V(F)$ the volume of the
fundamental domain of $H^3/\Ga$ which is defined in the Appendix A.

As discussed in the Introduction, the determinant of an elliptic differential operator  
requires a regularisation. One of the most 
used regularisation is the $\zeta$-function regularisation 
\cite{ray71-7-145,dowk76-13-3224,hawk77-55-133}. By means of it one has
\beq
\ln\det\fr{L_4}{\mu^2}=-\ze'\at0|\fr{L_4}{\mu^2}\ct
\:,\label{haw}\eeq
where the $\ze'$ is the derivative with respect to $s$ of 
the $\zeta$-function. In the standard cases,
the $\zeta$-function at $s=0$ is well defined and so by means of the 
latter formula one gets a finite result.  

It is well known that the meromorphic structure of the analytically continued 
$\zeta$-function, as well as the ultraviolet divergences of the one-loop 
effective action, can be related to the asymptotic properties 
of the heat-kernel trace, which in our case reads 
\beq
\Tr\exp{(-tL_4)}=\frac{\ell }{\sqrt{4\pi t}}\Tr\exp {(-tL_3)}\,.   
\label{8}
\eeq
For the rank-1 symmetric space $H^3/\Ga$ 
(the group $\Ga$ contains the identity, hyperbolic and parabolic elements) 
the trace of the operator $\exp{-(t L_3)}$ may be computed by using 
Eq.~(\ref{stf}) with the choice 
$h(r)=\exp\aq-t(r^2+\de^2)\cq$ and $\de^2=|\ka|+\al^2$
(here $\ka=R/6$ is the (negative) constant curvature of $H^3$).
We have
\beq
g(u)=\frac{e^{-t\de^2}e^{-u^2/4t}}{\sqrt{4\pi t}}\,,\hs 
g(0)=\frac{e^{-t\de^2}}{\sqrt{4\pi t}}\,,\hs 
h(0)=e^{-t \de^2}\,.
\label{g}\eeq
As a result, directly using Eq.~(\ref{fdiz})
\beq
\Tr e^{-t L_3}&=&e^{-t \de^2}\aq
\tilde P_1\ln t\:t^{-1/2}+\tilde A_0t^{-3/2}+\tilde A_1t^{-1/2}
-\frac{t}{\pi i}\int_{-\ii}^{\ii}
e^{-tr^2}f(\fr{ir}2)r\:dr\cq \nn\\
&&\hs\hs+\frac{e^{-t\de^2}}{\sqrt{4\pi t}}
\sum_{\ga}\sum_{k=1}^\ii
\frac{\chi(\ga)}{S_3(\ga,k)}
\:exp\at-\frac{k^2l_\ga^2}{4t}\ct
\:,\label{Kstf}\eeq
where
\beq
\tilde P_1=\frac{\sqrt{|\ka|}}{8\sqrt{\pi}}\:,\hs
\tilde A_0=\frac{V(F)}{(4\pi)^{3/2}}\,,\hs  
\tilde A_1=\frac{\sqrt{|\ka|}}{\sqrt{4\pi}}
\at C+\ln2+\frac\ga4\ct
\:,\label{}\eeq
$C$ being a computable constant (see Appendix B) 
and $\ga$ being the Euler constant.
The asymptotic behaviour of the last integral for $t\to0$ 
can be evaluated by using Eqs. ~(\ref{Kstf}) 
and (\ref{fdizExp}) and noting that the last contribution, 
namely the one associated with hyperbolic elements, is exponentially small
(for more details we refer the reader to Ref.~\cite{byts96-266-1}).
Besides the usual terms one has for the heat 
kernel in 3-dimensions, one gets terms with logarithmic factors
due to the presence of parabolic elements in $\Ga$. 
These terms are absent for co-compact group 
$\Ga$ (compact hyperbolic manifolds). 
The asymptotic for the heat kernel reads
\beq
\Tr e^{-tL_3}&\simeq& e^{-t\de^2}
\aq\tilde P_1\ln t\:t^{-1/2}+\tilde A_0t^{-3/2}+\tilde A_1t^{-1/2}
+\sum_{n=2}^\ii\tilde A_n t^{n-\frac32}\cq
\nn\\&\simeq&
\sum_{r=0}^{\ii}(A_r+P_r\ln t)\:t^{r-\frac32}
\:,\label{KP}\eeq
where  $\tilde A_2=\frac{1}{6\sqrt{|\ka|\pi}}$ and all 
the other coefficients are in priciple computable. We have put
\beq
A_r=\sum_{n=0}^{r}\frac{(-1)^n\tilde A_{r-n}\de^{2n}}{n!}\:,\hs
P_0=0\:,\hs 
P_r=\frac{(-1)^{r-1}\:\sqrt{|\ka|}
\:\de^{2(r-1)}}{8\sqrt{\pi}(r-1)!}
\:.\label{l3}\eeq
One has a similar short $t$-expansion for the heat-kernel trace related to 
the operator $L_4$. From Eqs.~(\ref{8}) and ~(\ref{KP}) one has
\beq
\Tr e^{-t L_4}\simeq  
\frac\ell{\sqrt{4\pi}}
\sum_{r=0}^{\ii}(A_r+P_r\ln t)\:t^{r-2}
\:.\label{te}\eeq

Let us analyse the consequences of the presence of logarithmic terms 
in the expansion Eq.~(\ref{te}). 
As usual, we may introduce the 
$\ze$-function associated with the elliptic operator $L_4$  
by means of the Mellin transform 
\beq
\ze(s|\fr{L_4}{\mu^2})=\frac1{\Ga(s)}\int_0^\ii dt\:t^{s-1}
\Tr e^{-t\frac{L_4}{\mu^2}}=\mu^{2s} \ze(s|L_4)
\:,\label{zf}\eeq 
valid for $\Re s>2$. In order to get the meromorphic structure of the 
function (\ref{zf}), we 
splits the integration range in the two intervals $[0,1)$ and 
$[1,\ii)$ obtaining in this way two integrals. 
The last one is regular for $s\to0$, while the behaviour of the
first one can be estimated by using the asymptotics, Eq.~(\ref{te}).
We have
\beq
\ze(s|L_4)\sim\frac{\ell}{\sqrt{4\pi}\Ga(s)}
\sum_{r=0}^\ii\aq\frac{A_r}{s+r-2}-\frac{P_r}{(s+r-2)^2} 
\cq\eeq
and so the meromorphic structure reads
\beq
\ze(s|L_4)&=&\frac\ell{\sqrt{4\pi}}\aq
\frac{A_0}{s-2}+\frac{A_1}{s-1}
-\frac{P_1}{(s-1)^2}-\frac{P_2}{s}\cq
\nn\\&&\hs\hs\hs
-\frac\ell{\sqrt{4\pi}}
\sum_{n=1}^{\ii}\frac{(-1)^n\:n!\:P_{n+2}}{s+n}
+\frac{\ell A_2}{\sqrt{4\pi}}+J_4(s)\,,
\label{ac}\eeq
where $J_4(s)$ is an analytic function. We see that in contrast with the 
usual cases, the analytic continuation of $\zeta$-function
is not regular at $s=0$, but it has a simple pole. It has to be noted 
also the presence of a double pole at $s=1$ and new simple poles
at $s=-n$ ($n\geq2$).

\s{The one-loop effective action}

We have seen in the previous Section that 
$\ze$-function regularisation cannot be used without modifications. In 
order to understand how to proceed, we shall work within the so called 
proper-time regularisation of one-loop effective action. More 
specifically, we define the regularised determinant by means of 
\cite{ball89-182-1,byts96-266-1} 
\beq
\at\ln\det L_4\ct_{\rho_\ep}
=-\int_0^\ii dt\:t^{-1}\rho_\ep(t)\Tr e^{-t L_4}
\:,\label{sr}\eeq 
in which $\rho_\ep(t)$ is a regularisation function necessary to deal 
with the singularities of $\Tr e^{-t L_4}$ present for $t\to0$. 
The properties of $\rho_\ep(t)$ are the following 
\cite{ball89-182-1,cogn93-48-790}:
\beq
\lim_{\ep\to0}\rho_\ep(t)=1\,,\hs
\lim_{t\to0}\rho_\ep(t)\sim t^{2+\de}\:,\hs\de>0\:.
\eeq
Of course, the latter requirement is valid for a sufficiently large 
$\ep$.
The $\zeta$-function regularisation may be obtained with the choice 
$\rho_\ep^{H}(t)=\frac{d}{d \ep}\at\frac{t^\ep}{\Ga(\ep)}\ct$. 
In fact, for a suitable $\Re\ep>0$ we have
\beq
\at\ln\det L_4\ct_H=-\int_0^\ii dt\:t^{-1}\frac{d}{d \ep} 
\at\frac{t^\ep}{\Ga(\ep)} \ct \Tr e^{-t L_4}=-\frac{d}{d \ep}
\ze(\ep|L_4)
\:.\label{z1}\eeq 
In the usual cases, namely when $\ze(s|L_4)$ is regular at $s=0$, 
the $\ze$-function regularisation directly gives the finite part of the 
effective potential. On the contrary, the other 
regularisations, besides a finite part, give divergent terms
proportional to the spectral coefficients (ultraviolet divergences). 
For example, the cutoff proper-time regularisation,  
corresponding to the choice $\rho_\ep^{C}(t)=\theta(t-\ep)$,
gives three ultraviolet divergent terms, proportional to
$\ep^{-2},\ep^{-1},\ln\ep$ respectively, while the
choice $\rho_\ep^{D}(t)=t^\ep$ \cite{dowk76-13-3224} 
(related to dimensional regularisation)
gives only a divergent term $-\ep^{-1}\ze(0|L_4)$.

As shown in the previous section, Eq.~(\ref{ac}), 
the $\zeta$-function for our case is not regular at the point $s=0$.
This means that $\zeta$-function regularisation for the
effective potential gives divergent contributions as well as 
all other regularisations. From Eqs.~(\ref{8}), ~(\ref{Kstf}) and 
\beq
\ze(s|\fr{L_4}{\mu^2}) &=& 
\frac{\ell}{\sqrt{4\pi}}\aq
\frac{\tilde A_0\de^4}{(s-1)(s-2)}
+\frac{\tilde A_1\de^2}{s-1}+\tilde A_2
\cq\at\frac{\de^2}{\mu^2}\ct^{-s}
\nn\\&&\hs\hs
+\frac{\ell\tilde P_1\de^2}{\sqrt{4\pi(s-1)}}
\aq\psi(s-1)-\ln\frac{\de^2}{|\ka|}\cq
\at\frac{\de^2}{\mu^2}\ct^{-s}
+\frac{G(s)}{\Ga(s)}\,,
\eeq
where $\psi$ is the logarithmic derivative of $\Ga$ and 
$G(s)$ is a function, regular in a 
neighbourhood of the point $s=0$, which  may be expressed in terms of the 
function $f(z)$, introduced in the Appendix B and defined 
by Eq.~(\ref{fdiz}) and the Selberg zeta-function $Z(s)$ related to the 
hyperbolic contribution (see for example \cite{byts96-266-1}). 
In particular, its value at the point $s=0$ is given by
\beq
G(0)&=&\frac{\ell\de^2}{\pi}\int_1^\ii
\at|\ka|r^2-1\ct^{1/2}\:\frac{Z'}{Z}(\de r+1)\:dr\nn\\
&&+\frac{\ell}{2\pi}
\int_{0}^{\ii}\at r^2+\de^2\ct^{-\frac12}\:
\aq irf(\fr{ir}2)-\fr{\sqrt{|\ka|}}6\cq\:dr\:\:\: +\:\:\: c.c.
\label{grad}\eeq
and its computation may require a numerical calculation.
As a result
\beq
\at\ln\det\fr{L_4}{\mu^2}\ct_H&=&
-\frac{\ell P_2}{\sqrt{4\pi}\ep^2}
+\frac{\ell P_2}{\sqrt{4\pi}}
\aq\frac12\at\ln\frac{\de^2}{\mu^2}\ct^2
+(\ga-1+\ln\frac{\mu^2}{|\ka|})
\at\ln\frac{\de^2}{\mu^2}-1\ct+\frac{\pi^2}6\cq
\nn\\&&\hs\hs\hs
+\frac{\ell}{\sqrt{4\pi}}
\aq A_2\ln\frac{\de^2}{\mu^2}
+\frac14A_0\de^4+A_1\de^2\cq-G(0)+O(\ep)
\:.\label{z11}\eeq

The structure of the ultraviolet divergences may be put in evidence
by using, for example, the cutoff proper-time regularisation. 
A straightforward computation gives
\beq
\at\ln\det\fr{L_4}{\mu^2}\ct_C
&=&\at\ln\det\fr{L_4}{\mu^2}\ct_H
+\ga\frac{\ell\tilde A_2}{\sqrt{4\pi}}
-\frac{\ell P_2}{\sqrt{4\pi}}
\aq \ga\ln\frac{\mu^2}{|\ka|}
+\frac{\ga^2}2-\frac{\pi^2}{12}\cq
\nn\\&&\hs
+\frac{\ell P_2}{\sqrt{4\pi}}
\aq\frac1{\ep^2}+\frac{\mu^2}{\de^2\ep}\at\ln\ep
+1-\ln\frac{\mu^2}{|\ka|}\ct+\frac12(\ln\ep)^2
+\ln\ep\ln\frac{\mu^2}{|\ka|}\cq
\nn\\&&\hs\hs\hs\hs
-\frac{\ell}{\sqrt{4\pi}}
\aq\frac{A_0\mu^4}{2\ep^2}
+\frac{A_1\mu^2}\ep-A_2\ln\ep\cq
\label{z22}\eeq 
which, apart divergent and finite terms, is equal to Eq.~(\ref{z11}).

We conclude this section with some remarks. 
First one can see that in general the $\zeta$-function regularisation 
is on the same level of the other proper-time regularisations and needs 
a divergent counterterm. If one makes use of another regularisation
new additional divergences appear. 
Such new ultraviolet divergences depend on the coefficient 
\beq
P_2=-\frac{\sqrt{|\ka|}}{8\sqrt{\pi}}
\aq m^2+\frac{\la\phi_c^2}{2}+|R|\at\fr16-\xi\ct\cq
\:.\label{dkir}\eeq
However, their dependence on $\phi_c$ is such that the usual one-loop 
renormalisation procedure may be applied. 
Furthermore, in the case of massless free scalar field with 
the conformal coupling $\xi=1/6$, we  have 
$P_2=0$ and thus, in this particular case the usual evaluation of 
the effective action by means of $\zeta$-function 
methods works without any modification. 

With regard to the general case, we will not discuss here the 
renormalization, but we limit ourselves to the following 
considerations. Since there exist a general proof that scale dependence of 
the effective potential, as well as the renormalization group 
equations, are standard in an arbitry curved space-time
\cite{buch92b}, one should expect that the additional divergence does 
not modify the standard renormalization procedure.    
 
\s{Finite temperature effects}
Since we have an ultrastatic space-time, the finite temperature 
effects can be studied considering the scalar field on 
$S^1\times H^3/\Ga$ 
satisfying periodic boundary conditions on the imaginary time, the 
period $\be$ being interpreted as inverse equilibrium temperature. 
Making use of the $\zeta$-regularisation, the logarithmic of the  
partition function is 
\beq
\ln Z_\be(\ep)=-\frac12\frac{d}{d\ep}\ag \frac{1}{\Ga(\ep)} 
\int_0^\ii dt\: t^{\ep-1} \Tr e^{-tL_\be/\mu^2} \cg
\:,\label{part}\eeq
where by $L_\be$ we indicate the Laplace-like operator acting on 
periodic scalar fields in $S^1\times H^3/\Ga$, while  $L_4$ acts on
fields in $\R\times H^3/\Ga$.
One easily has
\beq
\Tr e^{-t L_\be}=\frac{\be}{\sqrt{4\pi t}}\at 1+2\sum_{n=1}^\ii 
e^{-\frac{n^2\be^2}{4t}}\ct  \Tr e^{-t L_3}
=\at 1+2\sum_{n=1}^\ii 
e^{-\frac{n^2\be^2}{4t}}\ct\Tr e^{-t L_4} 
\:.\label{btr}\eeq 
Here we have put
\beq
\Tr e^{-t L_4}=\frac{\be}{\sqrt{4\pi t}}\Tr e^{-t L_3}
\:,\eeq
as in the zero temperature case with $\ell\to\be$.
As a result we have
\beq
\ln Z_\be(\ep)=-\fr12\ze'(\ep|\fr{L_4}{\mu^2}) 
-\int_0^\ii dt 
t^{-1}\sum_{n=1}^\ii e^{-\frac{n^2\be^2}{4t}} 
\Tr e^{-t L_4}+O(\ep)
\:,\label{ft}\eeq
in which the first contribution, depending linearly on $\be$, is related to the vacuum energy 
(effective potential)  and the second one is the statistical 
sum contribution. It is possible to show that this last contribution 
is ultraviolet finite, thus we may take the limit  $\ep\to0$. 

By using the Mellin-Parseval identity, the free energy can be conveniently 
written in the form \cite{byts96-266-1}
\beq
F_\be(\ep)=\frac1{2\be}\ze'(\ep|\fr{L_4}{\mu^2})
+\frac1{2\pi i\be}\int_{\Re z=c} 
\Ga(z)\ze(z|L_4)\at \fr{\be}{2} 
\ct^{-2z}\Ga(z)\ze_R(2z) \:dz 
\:,\label{mb}\eeq
where $\ze_R(z)$ is the Riemann $\ze$-function and $c>2$.
The integrand function has double poles at the points  
$z=-1,-2,-3,...$ and $z=1$, 
simple poles at $z=\frac12$ and $z=2$ and finally
a pole of order 3 at $z=0$.
Shifting the vertical contour to the left, 
with the help of the theorem of residues we obtain a small $\be$ 
expansion for the statistical sum $\hat F_\be=F_\be(\ep)-F_\ii(\ep)$.
It reads
\beq
\hat F_\be&=&\frac{V(F)\pi^2}{90\be^4}+
\frac{\pi^{\frac32}}{3\be^2}\aq 
A_1+2P_1\at\ga+\ln\frac\be2-\frac{6\ze_R'(2)}{\pi^2}\ct\cq
+\frac{\ze'(0|L_3)}{2\be}
\nn\\&&\hs\hs
+\frac{P_2}{\sqrt{4\pi}}\aq A_2
+2P_2\at\ga+2\ze_R'(0)+\frac12\ln\be\ct\cq\ln\be+...
\label{ht}\eeq
It should be noted that the presence of 
parabolic elements in the discrete group of isometry leads to the 
appearance of new terms in the high temperature expansion, only the 
leading (Planckian) one being left unchanged.

\s{Conclusions}
In this paper we have considered a self-interacting scalar field in 
an ultrastatic space-time whose spatial section is a non-compact 
hyperbolic manifold with finite volume. By means of the Selberg trace 
formula we have investigated the one-loop effective action. 
It has been shown that the asymptotics of the heat-kernel trace related to the small disturbance 
operator, a Laplace-type operator acting on the manifold 
$R\times H^3/\Ga$, 
contains logarithmic terms in $t$, absent in the smooth or compact 
background. 
As a consequence, we have seen that the $\ze$-function can be analytically 
continued and its  meromorphic structure admits a simple pole at $s=0$. 
In order to analyse this new situation,  a general class 
of regularisations (proper-time regularisations) have been considered 
and we have observed that  also the $\ze$-function 
regularisation of the one-loop 
effective action  requires a divergent counterterm, 
as well as all other regularisations. We have noticed, however, that 
no problem arises as soon as one is dealing with a conformally coupled 
free massless scalar field. 

The finite temperature effects have also been investigated and a 
general expression for the 
one-loop free energy at temperature $\be^{-1}$ has been presented. 
It has to be noticed that new terms appear in the high-temperature expansion. 
As a consequence, the quantum corrections to the mass, which are 
defined by means of the relation
\beq
V(\phi_c)=\La_{eff}+\frac{1}{2} 
\aq m^2+m^2(\be)\cq \phi_c^2+ 
O(\phi_c^4)
\:,\label{cm}\eeq
receive new contributions, which may play a role 
in the discussion of the phase 
transition of the system, that is in the problem of symmetry breaking 
(or restoration) in some cosmological scenarios.

Another potentially interesting application of the model we have 
studied deals with the problem of 
the divergences one encounters in the evaluation of the entropy (the 
entanglement entropy) for a 
scalar field in Rindler space-time and in the related mechanical 
statistical entropy of the black hole 
\cite{thoo85-256-727,bomb86-34-373,call94-333-55}. This issue has been 
recently discussed in Ref.~\cite{frol96-54-2711}.  
 It is a common feature of the all methods used in the analysis of the 
mechanical-statistical entropy, namely the entropy of quantum fields 
in a background with a horizon, the fact that the density of states is 
blowing up near the horizon. 
It has been shown that the leading divergence is
associated to the divergence of the spatial optical volume. 
For Rindler space-time, the 4-dimensional conformal optical metric 
is ultrastatic and the  spatial optical 
section turns out to be locally the 
hyperbolic 3-dimensional space $H^3$, which obviously, 
being non-compact, has an infinite volume. 
For physical reasons, the spatial 
metric must be non-compact. However, we remind that, besides 
the huge number of 
compact hyperbolic manifolds (the ones whose discrete group of 
isometries contain hyperbolic and elliptic elements), 
there exist also non-compact hyperbolic manifolds with finite volume. 

With regard to this issue, it might be interesting to consider the 
space-time we are dealing with, i.e.  $ R \times H^3/\Ga $ 
as the conformal optical counterpart 
of a space-time which (at least locally) 
is the Rindler one with some identifications. 
As has been considered in this paper, the discrete group $\Ga$
has the form $SL(2,\Z+i\Z)/\{\pm Id\}$, generated by parabolic mappings 
and related to non-compact
manifold with invariant fundamental domain of finite volume. 
In the Appendix A a homeomorphism
between $H^3/\Ga$ and boundaryless manifold 
$(S^1\times U_c)/\partial M$ has been presented. The
manifold $(S^1\times U_c)/\partial M$ can be constructed by means  of 
the compactification of two
planar coordinates of a Rindler-like spatial section and 
the identification of the faces of the
corresponding polyhedron. In some sense, this compactification procedure 
could be similar to the regularisation with
the help of volume cutoff parameter introduced in 
Ref.~\cite{frol96-54-2711}.

\ack{We thank K.~Kirsten and L.~Vanzo for useful discussions. A.A.B.~would like 
to thank I.N.F.N. for financial support and Prof.~M.~Toller for kind 
hospitality at Department of Physics, University of Trento.  
The research of A.A.B.~was supported in part 
by Russian Foundation for Fundamental Research
grant No.~95-02-03568-a and by  
Russian Universities grant No.~95-0-6.4-1.}
\appendix

\s{Fundamental domain of the discrete group $SL(2,\Z+i\Z)/\{\pm Id\}$}

In this first Appendix, we will summarise the geometry and local isometry 
associated with a simple 3-dimensional complex Lie group. We shall 
consider discrete subgroups $\Ga\subset SL(2,\C)/\{\pm Id\}$, where  
$Id$ is the $2\X2$ identity matrix and is an isolated element of the 
$\Ga$. The group $\Ga$ acts discontinuously at 
the point $z\in\bar\C$, $\bar\C$ being the extended complex plane.
We recall that a transformation $\ga\neq Id$ $\ga\in\Ga$, with
\beq
\ga z=\frac{az+b}{cz+d}\,,\hs ad-bc=1\,, \hs a,b,c,d\in\C,
\eeq 
is called elliptic if $(\Tr\ga)^2=(a+d)^2$ satisfies 
$0\leq(\Tr\ga)^2 < 4$, hyperbolic if $0\leq(\Tr\ga)^2 > 4$, 
parabolic if $(\Tr\ga)^2=4$ and loxodromic 
if $(\Tr\ga)^2\in\C\backslash\aq0,4\cq$. The classification of these 
transformations can also be based on the properties of their fixed 
points, the number of which is one for the parabolic transformations 
and 2 for the other cases.

The element $\ga\in SL(2,\C)$ acts on $z=(y,w)\in H^3$
(as usual here we choose units in which $\ka=-1$), 
$w=x_1+ix_2$ by means of the following 
linear-fractional transformation:
\beq
\ga z=\at\frac{y}{|cw+d|^2+|c|^2y^2}\,,\frac{(aw+b)(\bar c \bar 
w+\bar d)+a \bar c y^2}{|cw+d|^2+|c|^2y^2} \ct 
\:.\eeq 

The isometric circle of a transformation 
$\ga\in SL(2,\C)/\{\pm Id\}$ 
for which $\ii$ is not a fixed point is defined to be the circle 
\beq
I(\ga)=\{z:\:\:\:|\ga z|=1\}\:,\hs\mbox{ or}\hs
I(\ga)=\{z:\:\:\:|z+d/c|=|c|^{-1}\}\:,\:\:\: c\neq0\:.
\eeq
A transformation $\ga$ carries its isometric circle 
$I(\ga)$ into $I(\ga^{-1})$.

The isometric fundamental domain of a Fuchsian group (Kleinian group 
without loxodromic elements) has the following  structure: it is bounded by 
arcs of circles orthogonal to the invariant circle and consists either 
of two symmetric components or of a single component, while the 
mappings connecting its equivalent sides, generate the whole group. In 
many cases, it is more convenient to deal with other fundamental 
regions. For example, the so-called normal fundamental Dirichlet polygons  
are often used for Fuchsian groups and we will follow this approach 
here.

Now we consider a discrete group $\Ga$ of a special kind.
Let $G=SL(2,\C)/\{\pm Id\}$, then for $\Ga\subset G$, one can choose
$\Ga=SL(2,\Z+i\Z)/\{\pm Id\}$, where $\Z$ is the ring of integer 
numbers. The element $\ga\in\Ga$ will be identified with 
$-\ga$. The group $\Ga$ has, within a conjugation, one maximal 
parabolic subgroup $\Ga_\ii$ ($c=0$). Thus, the fundamental domain 
related to $\Ga$ has one parabolic vertex and can be taken in the form
\cite{venk73-125-3,lang85b}
\beq
F(\Ga)=\ag (y,w): x_1^2+ x_2^2 +y^2 >1,\:\:\:  
-\fr12<x_2<x_1<\fr12\cg\:.\label{a2}\eeq  
As an example let us consider the parabolic mappings
\beq
g_1(z)=z+1\,,\hs  g_2(z)=z+i 
\:.\eeq
If we identify the faces of the polyhedron, Eq.~(\ref{a2}), we get a 
manifold $M(\Ga)$ that is homeomorphic to a punctured torus 
$S^1\otimes S^1\otimes\aq-\frac{1}{2},\frac{1}{2}\ct=U_c\otimes S^1$, 
where $U_c=\ag z:\:\:\:0<|z|\leq\frac{1}{2}\cg$ is a punctured cylinder. 
It is turned into a hyperbolic manifold by removing the boundary 
$\partial M(\Ga)$, 
which is homeomorphic to the torus $S^1\otimes S^1$.

\s{A contribution to the Selberg trace formula associated with the 
cusp form}
               
Here we will discuss the Selberg trace formula on the 
space $H^3$, which can be constructed as an expansion in eigenfunctions of 
the automorphic Laplacian. To begin with, we assume that the group 
$\Ga$ is generated by parabolic mappings. Since the discrete group 
$\Ga$ has a cusp at $\ii$ ($c=0$), each element of the stabiliser 
$\Ga_{\ii}$ is a translation, as we have seen in the Appendix A. 
We compute the conjugacy class 
$\ag\ga\cg_{\Ga}$, $\ga\in\Ga_{\ii}$ 
with $\ga$ different from identity.
If 
\beq
\ga=\left(\matrix{1& n_1+in_2\cr0&1\cr}\right) 
\:,\hs n_1,n_2\in\Z \:,
\label{b1}\eeq 
then the conjugacy class with representative $\ga$ consists in 
$\ga$  and $\ga^{-1}$, where
\beq
\ga^{-1}=\left(\matrix{1&-n_1-in_2\cr0&1\cr} \right)
\:.\label{b2}
\eeq 
The remaining conjugacy classes have the representatives in
$\Ga_{\ii}$ of the form
\beq
\ga_1=\left(\matrix{i&0\cr0& -i\cr}\right)\:,\,\,\,
\ga_2=\left(\matrix{i&1\cr0& -i\cr}\right)\:,\,\,\,
\ga_3=\left(\matrix{i&-i\cr0&-i\cr}\right)\:,\,\,\,
\ga_4=\left(\matrix{i&1-i\cr0&-i\cr}\right)\:.
\eeq 
The centralisers related to these representations read
\beq
\Ga^\ga=\left(\matrix{1&m_1+im_2\cr0&1\cr} \right) 
\:,\hs  m_1,m_2\in \Z\:,
\eeq 
\beq
\Ga^1=\Ga^{\ga_1}=\ag
\left(\matrix{1&0\cr0&1\cr}\right)\:,
\left(\matrix{i&1\cr0&-i\cr}\right)\:,
\left(\matrix{0&1\cr -1&0\cr}\right)\:,
\left(\matrix{0&i\cr i&0\cr}\right)
\cg\:,\nn\eeq
\beq
\Ga^2=\Ga^{\ga_2}=\ag
\left(\matrix{1&0\cr0&1\cr}\right)\:,
\left(\matrix{i&1\cr0&-i\cr}\right)\:,
\left(\matrix{i&0\cr2& -i\cr}\right)\:,
\left(\matrix{-1&i\cr 2i&1\cr}\right)
\cg\:,\nn\eeq
\beq
\Ga^3=\Ga^{\ga_3}=\ag
\left(\matrix{1&0\cr0&1\cr}\right)\:,
\left(\matrix{i&-i\cr0&-i\cr}\right)\:,
\left(\matrix{1&-1\cr2&-1\cr}\right)\:,
\left(\matrix{i&0\cr2i&-i\cr}\right)
\cg\:,\nn\eeq
\beq
\Ga^4=\Ga^{\ga_4}=\ag
\left(\matrix{1&0\cr0&1\cr}\right)\:,
\left(\matrix{i&1-i\cr0&-i\cr}\right)\:,
\left(\matrix{i&0\cr1+i&-i\cr}\right)\:,
\left(\matrix{1&-1-i\cr1-i&-1\cr}\right)
\cg\:.\nn\eeq

Let us consider an arbitrary integral operator with kernel $k(z,z')$.
Invariance of the operator is equivalent to fulfillment of the condition 
$k(\ga z,\ga z')=k(z,z')$ for any $z,z'\in H^3$ and 
$\ga\in PSL(2,\C)$. So the kernel of the invariant operator is a 
function of the geodesic distance between $z$ and $z'$. 
It is convenient to replace such a distance with the fundamental 
invariant of a pair of points  
$u(z,z')=|z-z'|^2/yy'$, thus  $k(z,z')=k(u(z,z'))$ . Let $\la_i$ be 
the isolated eigenvalues of the self-adjoint extension of the Laplace operator 
and let us introduce a suitable analytic function $h(r)$ and 
$r^2_j=\la_j-1$. It can be shown that $ h(r)$ 
is related to the quantity $k(u( z,\ga z))$
by means of the Selberg transform. Let us denote by $g(u)$ the 
Fourier transform of $ h(r)$, namely
\beq
g(u)=\frac{1}{2\pi}\int_{-\ii}^\ii e^{-iru} h(r) dr
\:.\eeq

For one parabolic vertex  let us introduce a subdomain $F_Y$ of the fundamental 
region $F(\Ga)$ by means
\beq
F_Y=\ag z \in F(\Ga), z=\ag y, \vec x \cg | y \leq Y \cg
\:,\label{b7}\eeq
where $ Y$ is a sufficiently large positive number. Let us suppose 
that $h(r)$ is a even analytic function in the strip 
$|\Im r|<1+\ep$ ($\ep>0$) and  $h(r)=O(1+|r|^2)^{-2}$. 
Then for $N=3$ the following formula holds \cite{venk73-125-3}:
\beq
\sum_j h(r_j)&=&\lim_{Y\to\ii}\ag
\int_{F_Y}\sum_{\ga\in\Ga}k(u(z,\ga z))\:d\mu(z)
\cp \nn\\
&&\hs\hs-\ap
\frac{1}{2\pi}\int_0^\ii h(r)\int_{F_Y}|E(z,1+ir)|^2\:d\mu(z)dr 
\cg\:,\label{b8}\eeq 
in which $d\mu(z)=y^{-3}dydx_1dx_2$ is the invariant measure on $H^3$
and $E(z,s)$ is the 
Eisenstein-Maass series associated with one cusp, namely
\beq
E(z,s)=\sum_{\ga\in(\Ga/\Ga_\ii)} y^s(\ga z)\,,\hs x_2(z)=\Im z
\:.\label{a8}\eeq
The series (\ref{a8}) converges absolutely for $\Re s>1$ and uniformly in $z$ 
on compact subset of $H^3$. All poles of $E(z,s)$ are contained in 
the union of the region $\Re s<1/2$ and the interval 
$\aq1/2,1\cq$ and those contained in such an interval are simple.
Furthermore, for each $s$, the series  $E(z,s)$ is a real 
analytic function on $H^3$, automorphic relative to the group $\Ga$ 
and satisfies the eigenvalues equation
\beq
\bar \lap_3 E(z,s)=s(s-1)E(z,s)
\:.\label{b9}\eeq
The asymptotic expansion of the second integral in Eq.~(\ref{b8}) can be 
found with the help of Maass-Selberg relation 
(for details see Ref.~\cite{venk73-125-3}). 
For $Y\to\ii$ one has the following result
\beq
\frac{1}{2\pi}\int_0^\ii h(r)\int_{F_Y}|E(z,1+ir)|^2\:d\mu(z)dr
&=&g(0)\ln Y+\frac{h(0)}{4}S(1)
\nn\\&&
-\frac{1}{4\pi}\int_{-\ii}^\ii  h(r) 
\frac{S'(1+ir)}{S(1+ir)}\:dr
+O(1)
\:.\label{b10}\eeq
The function $S(s)$ (in the general case it is the S-matrix) is given 
by a generalised Dirichlet series, convergent for $\Re s>1$,
\beq
S(s)=\frac{\sqrt\pi\Ga(s-\frac{1}{2})}{\Ga(s)}
\sum_{c\neq0}\sum_{0\leq d<|c|} |c|^{-2s}
\:,\label{b11}\eeq
in which the sums are taken over all pairs $c,d$ of the matrix 
$\at\matrix{*&*\cr c&d\cr}\ct\subset\Ga_\ii\backslash\Ga/\Ga_\ii$.
Also the poles of the meromorphic function $S(s)$ are contained 
in the region $\Re s<1/2$ and in the interval $\aq1/2,1\cq$. 
The functions $E(z,s)$ and $S(s)$ can 
be analytically extented on the whole complex 
s-plane, where they satisfy the functional equations
\beq
S(s)S(1-s)=Id\:,\hs 
\overline{S(s)}=S(\bar s)\:,\hs
E(z,s)&=&S(s)E(z,1-s)
\:.\label{b12}\eeq

It should be noted that the terms of the trace formula associated with 
the elements $\ga$ and $\ga^{-1}$ coincide. Then the contribution to 
the first integral of Eq.~(\ref{b8}), which comes from all conjugacy 
classes of the $\ga$-type 
($\ga\in\Ga^{\ga}$), for $Y\to\ii$ can be written as follows
\beq
\int_{F_Y}\sum_{\ga\in\Ga}k(u(z,\ga z))\:d\mu(z)
&=&\at \ln Y+C\ct g(0)+\frac{h(0)}{4} 
\nn\\&&\hs
-\frac{1}{4\pi}\int_{-\ii}^\ii h(r)
\psi(1+\fr{ir}{2})\:dr+O(1)
\:,\label{b13}\eeq
where $\psi$ is the logarithmic derivative of the Euler $\Ga$-function
and $C$ a computable constant which reads \cite{venk73-125-3}
\beq
 C&=&\frac{5\ln2}{16}-\frac\ga2+C_0\:,\nn\\
C_0&=&\lim_{N\to\ii}\:\:\frac1{4\pi}
\sum_{i=1}^{N}\aq|\xi^i|^{-2}-2\pi\ln\frac{|\xi^{i+1}|}{|\xi^i|}\cq
-\frac12\ln|\xi^1|
\:.\eeq
In the latter equation $\ga$ is the Euler-Mascheroni constant
and $\xi$ is a two-dimensional vector, such that
$\ga z=\{y,\om+\xi\}$, $\xi\neq0$, $|\xi^{i+1}|\geq|\xi^i|$.

For the derivation of the Selberg trace formula, 
which we have used in Sec. 3, one has to consider the contributions 
coming from the identity element in $\Ga$, the normalised 
Eisenstein-Maass series, Eq.~(\ref{a8}) 
and all  $\ga$-type conjugacy classes, 
Eq.~(\ref{b13}).  The final result reads
\beq
\sum_j h(r_j)&-&\frac{1}{4\pi}\int_{-\ii}^\ii h(r) 
\frac{S'(1+ir)}{S(1+ir)}\:dr+\frac{h(0)}{4}S(1)
\nn\\ &=&
V(F)\int_0^\ii\frac{r^2}{2\pi^2}\:h(r)\:dr
+\sum_{\ga}\sum_{k=1}^\ii\frac{\chi(\ga)}{S_3(\ga,k)}g(kl_\ga)
\nn\\&&\hs
+Cg(0)+\frac{h(0)}{4}
-\frac{1}{4\pi}\int_{-\ii}^\ii h(r)\psi(1+\fr{ir}{2})\:dr
\nn\\ &=&
V(F)\int_0^\ii\frac{r^2}{2\pi^2}\:h(r)\:dr
+\sum_{\ga}\sum_{k=1}^\ii\frac{\chi(\ga)}{S_3(\ga,k)}g(kl_\ga)+
(C+\frac12\ln2)g(0)
\nn\\&&\hs
-\frac1{2\pi}\int_{0}^{\ii}h(r)\:\ln r\:dr
+\frac{1}{2\pi i}\int_{-\ii}^\ii h'(r) f(\fr{ir}{2})\:dr
\:.\label{stf}\eeq
The first term in the r.h.s. of Eq.~(\ref{stf}) is the contribution of 
the identity element, $V(F)$ is the (finite) volume 
of the fundamental domain $F$ with respect to the measure $d\mu$ and 
the second term is the standard contribution of hyperbolic elements, 
which has been considered, for example, in Ref. \cite{byts96-266-1} 
and we refer to this reference for details. 
The function $f(z)$ is defined by (see Ref.~\cite{grad80b})
\beq
f(z)=\frac12\sum_{k=1}^{\ii}\frac{k}{(k+1)(k+2)}
\sum_{n=1}^{\ii}\frac1{(n+z)^{k+1}}
\label{fdiz}\eeq 
and has an aymptotic expansion for large $|z|$ in terms of
the Bernoulli numbers $B_k$. It reads
\beq
f(z)\sim\sum_{k=1}^{\ii}\frac{B_{2k}}{2k(2k-1)z^{2k-1}}
\:.\label{fdizExp}\eeq 
%%%%%%%%%%%%%%%%%%%%%%%%%%%%%%%%%%%%%%%%%%%%%%%%%%%%%%%%%%

\end{document}